\documentstyle[12pt]{article}
\def\ni{\noindent} 

\begin{document}
\begin{center}
{\Huge {\bf Axial Anomaly from the BPHZ regularized BV master equation}}\\
\end{center}
\vspace{1cm}

{\large Everton M. C. Abreu  and   Nelson R. F. Braga  \\
\vspace{1cm}

 Instituto de F\'\i sica, Universidade Federal  do Rio de Janeiro,\\
Caixa Postal 68528, 21945  Rio de Janeiro,
RJ, Brazil\\

\vspace{1cm}
\abstract
A BPHZ renormalized form for the master equation of the field antifield (or BV) quantization has recently been proposed by De Jonghe, Paris and Troost. 
This framework was shown to be very powerful in calculating gauge anomalies. 
We show here that  this equation can also be applied in order to  calculate a global anomaly  (anomalous divergence of a classically conserved Noether current), considering the case of QED. This way, the fundamental  result about the anomalous contribution to the Axial Ward identity  in standard QED (where there is no  gauge anomaly)  is reproduced in this BPHZ regularized BV framework.
\vfill\eject
\section{Introduction}
The Field Antifield, or Batalin Vilkovisky (BV) quantization\cite{BV,HT,GPS,DJ} is presently considered to be the most general procedure for the quantization 
of gauge field theories.  The calculation of gauge  anomalies at one loop order
in this framework, pioneered in \cite{TPN}, has lead to important results summarized, for example, in \cite{GPS}. 
However, the potentiality of the BV procedure in calculating higher loop order contributions has only very recently been exploited.
Two different approaches have been proposed in order to calculate higher loop
anomalies. One, presented in \cite{Pa}, is the incorporation in the BV context of the non local regularization procedure. This procedure was also 
shown in \cite{PT} to be useful in deriving consistency conditions at two loop order.
The second  approach\cite{DPT}, that will be discussed here, is based on the BPHZ renormalization scheme.  In contrast to  the  standard BV formulation,
one avoids the use of the  $\,\,\Delta\,\,$ operator, that leads to singular results. One rather writes out correlation functions, taking into account 
possible violations of the BRST invariance of the effective action, expressed by the Zinn Justin equation.  Then a BPHZ renormalized version for this correlation functions is considered. This allows the computation of higher loop gauge anomalies by following the systematic BPHZ  prescriptions.

Let us now turn to the case of global anomalies.
At the classical level, the invariance of the action with respect to  a global transformation correspond to the conservation of the associated Noether current. Quantum corrections may however change this picture. The vacuum expectation value of the divergency of a classically conserved current may be non vanishing\cite{ABJ,Ja}. In other words, correlation functions involving the divergence of a Noether current may have a non trivial behavior. This  is  what is called a global anomaly.         

In the standard Field Antifield quantization the generating functional does not reflect this possible important non trivial quantum behavior. However, it was recently shown in ref\cite{AB} that one can enlarge the symmetry content of some field theory (and correspondingly the associated field antifield space), by trivially gauging some Abelian global symmetry, in such a way that one is naturally forced  to introduce an extra  quantum correction to the action, in order to solve the master equation.  This correction generates the appropriate anomalous Ward identities involving the divergence of the associated  Noether current. The original theory is recovered if we fix the new symmetry at some particular gauge.  It was also shown in \cite{AB} that
this trivial gauging of abelian global symmetries does not change the cohomology, therefore no extra gauge anomalies are introduced.

The purpose of the present letter is to show that this procedure  for calculating global anomalies can also be applied to the BPHZ renormalized BV master equation. Although this equation was conceived for calculating gauge anomalies, we will show, considering the case of QED ( where there is no gauge anomaly)  that it can also reproduce the anomalous non conservation of the vacuum expectation value of a Noether (global) current ( the Axial current in this case).
In section (2) we will very briefly review some facts about the BPHZ renormalization scheme, that will be essential in our application.
In section (3) we present the Renormalized master equation and in section (4) 
we analyze the case of QED, showing how to incorporate the global anomaly.
Some concluding remarks are left for section (5).  

\section{BPHZ Renormalization}

The BPHZ renormalization scheme\cite{BPHZ} provides a systematic way of giving a well defined meaning to Greens functions of composite operators. Finite results are associated to any divergent integral by means of subtracting a Taylor series of the integrand in the external momenta.    
Even at higher loop  order,  where the Greens functions involve integrations in overlapping momenta, the so called Forest formula defines unambiguously a subtraction procedure that eliminates all the ultraviolet singularities.

The definition of a product of fields at the same point is in general  problematic  in Quantum Field Theory as it may involve divergent integrals. 
The BPHZ scheme provides a way of defining, given any composite operator ${\cal O }( x ) $, a sequence of normal products

\begin{equation}
N_\delta [ {\cal O }(x)  ] 
\end{equation}

\noindent (where $\delta\,$, the order of the normal product,  is an integer equal or greater than
the canonical dimension of the operator ${\cal O} (x)$) 
in such a way that the integrals that appear in the Gell-Mann Low expansion of the  Greens functions  are replaced by the corresponding subtracted ones. In the case of massless theories one needs to introduce a mass term in order to avoid infrared divergencies.
In order to relate normal products of different order one uses the so called Zimmermann identities

\begin{equation}
N_{_{\alpha 1}} [ {\cal O}(x) ] \,=\, 
N_{_{\alpha 2}} [ {\cal O}(x) ] \,+\, 
\sum \, r_i \, N_{_{\alpha i}} \,[\, {\cal O}_i (x) \,]
\end{equation}

The application of the BPHZ renormalization to  QED is widely
discussed in the literature.
However, it  will be important for us to write out a result that can be obtained from \cite{LS}, if one takes, after the renormalization computations, the zero mass limit:

\begin{equation}
\label{BAX}
\partial^\mu \, \langle 0 \vert\, T (\,\, N_3\, [\,\, J_{_{\mu\,5}}^{^{BPHZ}}\, ]\, \,X\,
\vert 0 \rangle\,=\,\, r\,  \langle 0 \,\vert\, T\, 
(\, N_4\, [ \, F_{\mu\nu}\,{\tilde F}^{\,\mu\nu}\,\,  ]\, X
\vert\, 0\, \rangle
\end{equation}

\ni where we are denoting as $\,\,J_{_{5\,\mu}}^{^{BPHZ}}\,\, $ the quantity that plays the role of Axial current in the BPHZ renormalized framework:

\begin{equation}
\label{BCURR}
J_{_{\mu\, 5}}^{^{BPHZ}}  \,=\,(1 + d_1 - s - 2\beta r )  \,
{ \overline \psi}
\gamma_\mu  \gamma^5 \psi \,=\,\alpha_1 \, {\overline \psi} \gamma_\mu 
\gamma^5 \psi
\end{equation}

\ni and $\,\,r = e^2/(4\pi)^2  $ and  $d_1 , s, \beta $ are constant coefficients defined in  \cite{LS}, that will not be relevant here.

\section{Anomalies in the BV-BPHZ framework} 

Considering the generator of connected Greens functions ${\cal W} [J^A\,,\,\Phi^{\ast\,A}\,]$, depending on the sources $J$  and the antifields $\Phi^{\ast}$ one can 
introduce the effective action in the usual way by:

\begin{equation}
\Gamma [\phi^A_{cl\,\,}\,,\,\phi^{\ast\,A}\,]\,=\,
{\cal W} [J^A\,,\,\phi^{\ast\,A}\,] \,-\,
J^A\,\phi^A_{cl\,}
\end{equation}

Let us summarize some results from ref\cite{DPT}. 
The most general behavior of the effective action with respect to BRST transformations is described by the Zinn Justin equation:

\begin{equation}
\label{ZJ}
{1\over 2} \{ \Gamma \,,\,\Gamma \} \,=\,
{\delta^L \Gamma \over \delta \phi^A_{cl\,\,} }
{\delta^R \Gamma \over \delta \phi^{\ast\,A} }
\,=\, -\,i\hbar \,(\,{\cal A} . \Gamma \, )
\end{equation}

\noindent where $(\,{\cal A} . \Gamma \, )$ is the generating functional of 
One particle irreducible diagrams with the insertion of the anomaly 
${\cal A} $.
Defining the generating functional as:

\begin{equation}
Z \,[\,J^A\,,\,\phi^{\ast\,A}\,] \,=\,\int\,{\cal D}\phi 
\,\,exp {i\over \hbar}\,\{ S[\phi\,,\,\phi^\ast \,] \,+\,J^A\,\phi^A \,\,\}
\end{equation}
 
\noindent and the expectation value of some operator $\,X\,$ in the presence of sources $J^A$ as

\begin{equation}
\langle \, X\,[\,\phi\,,\,\phi^\ast\,]\,\rangle_{_J}\,\,=\, 
\,\int\,{\cal D}\phi \,X\,
\,\,exp {i\over \hbar}\,\{ S[\phi\,,\,\phi^\ast \,] \,+\,J^A\,\phi^A \,\,\}
\end{equation}

One can show that equation (\ref{ZJ}) can be translated into a set of relations involving expectation values of derivatives of the quantum action $S$ with respect to the antifields. The BPHZ renormalization can than be introduced, and these relations reads:

\begin{equation}
\label{MasterB1}
\int d^n x J_A (x) \langle \, N_{d^A} \Big[ 
{\delta^L S \over \delta \phi^{\ast}_A (x)} \Big]\,
\rangle_{_J} 
\,=\, i\hbar \,\int d^n x \langle \, N_n [ {\cal A}(x) ]
\rangle_{_J}
\end{equation}

\noindent expanding this equation in the sources we find a relation that will be useful 
in our application.

\begin{eqnarray}
\label{MasterB2}
\int d^nx \langle \Big( N_{d^A} \Big[{\delta^L S \over \delta 
\phi^{\ast}_A (x) }
\Big]\, { \delta^L \over
\delta \phi^A (x)} (-1)^{\epsilon_A} \Big)
\phi^{A_1} (x_1)  \phi^{A_2} (x_2)\,...\, \phi^{A_n} (x_n)
\,\rangle_{_J} & &\nonumber \\
 = \,\,\int d^nx \langle N_n \Big[ {\cal A} (x)
\Big] 
\phi^{A_1} (x_1)  \phi^{A_2} (x_2)\,...\, \phi^{A_n} (x_n)
\,\,\rangle_{_J}& &\nonumber\\
\end{eqnarray}

\section{Axial anomaly in QED}
Let us consider QED in four dimensions, represented by the classical action:

\begin{equation}
\label{LQED}
S_0 = \int d^4 x \Big[ - {1\over 4}  F^{\mu\nu}\,
F_{\mu\nu} \, + \,i\overline \psi \gamma^\mu \big(
\partial_\mu -ig A_\mu \big) \psi \Big]
\end{equation}

This theory has no gauge anomaly. However, as is well known,
there is a global anomaly corresponding to the violation of the conservation of the Axial current. This fact has very important consequences, discussed for example in \cite{Ja}.
If we follow the procedure of \cite{AB} in order to investigate this global anomaly,  now in the context of the BV-BPHZ procedure, we must gauge the axial transformation. We introduce a scalar field $\rho$ and couple it with the BPHZ Axial current given by eq. (\ref{BCURR}),  adding to the
action (\ref{LQED}) the term

\begin{equation}
\label{EXTRA}
S_0^\prime \,=\,\int d^4x \,\,\, J_{_{5\,\mu}}^{^{BPHZ}} \,\partial_\mu \rho\,=\,  \int d^4x \,
 e \,\alpha_1 \,\overline\psi \gamma^\mu \gamma_5 \partial_\mu \rho\,\psi \,\,.
\end{equation}

\noindent The action $S_0 + S_0^\prime $ has, besides the standard (non axial) gauge invariance of QED, the additional local symmetry:

\begin{eqnarray}
\label{AXIAL}
\delta \psi (x) &=& ie \alpha_1 \lambda (x) \gamma_5 \psi (x)
\nonumber\\
\delta \overline\psi (x) &=&  ie \alpha_1 \lambda (x) \overline \psi (x) \gamma_5 \nonumber\\
\delta \rho &=& \lambda (x)\,\,.
\end{eqnarray}

\noindent and, as discussed in \cite{AB}, we recover the original theory if we fix the gauge corresponding to $\rho = 0$. It is also important to stress\cite{AB} that the new field $\rho$ and his corresponding ghost field will form a BRST doublet\cite{PS} and will therefore be absent from the cohomology. In other words, for a theory like QED, without gauge anomalies, the process of gauging the global axial symmetry will not render the theory anomalous. 

 Before writing the BV action at classical level 
associated to $S_0 + S_0^\prime$, it is convenient to perform a canonical transformation, as prescribed in
\cite{VV}:

\begin{eqnarray}
\rho &\rightarrow& -\,\omega^\ast\nonumber\\
\rho^\ast  &\rightarrow& \omega
\end{eqnarray}

\noindent This way we get the quantum BV action:

\begin{eqnarray}
S &=& \int d^4x \Big( -{1\over 4} F_{\mu\nu} F^{\mu\nu} + \overline\psi \gamma^\mu \big[ \partial_\mu
- ie A_\mu - i \alpha_1 \gamma_5 \partial_\mu \omega^\ast \big] \psi
\nonumber\\
&-& ie \psi^\ast ( c + \gamma_5 d ) \psi + 
ie {\overline\psi}^\ast ( c + \gamma_5 d ) \overline\psi +   A^\ast_\mu \partial^\mu c + \omega d \Big]\, \nonumber\\ &+& \,\hbar \, M_1 
\end{eqnarray}

\noindent where the ghosts $c$ and $d$ are associated respectively to the original gauge symmetry and to the new axial one ( eq. {\ref{AXIAL}) and $M_1 $ represents a possible one loop order correction to the action that, as we will see will have a non trivial role in generating the anomalous divergence of the Axial current.

As there is no gauge anomaly in QED, the righthand side of 
equations (\ref{MasterB1}) and (\ref{MasterB2}) vanishes
(${\cal A}\,=\,0$). 
It is important to stress that the so called Adler Bell Jackiw anomaly, corresponding to the righthand side of eq. (\ref{BAX}) is not a gauge anomaly. The transformation associated to $\,\,J_{\,5\,\mu}\,\,$ is a global symmetry of  (\ref{LQED}) and, as a consequence of the Noether theorem, this current is conserved at the classical level. What we call a global anomaly is a violation in this conservation law at quantum level.
In our procedure, introducing the extra term (\ref{EXTRA}) in the action, we gauge this symmetry but, as in \cite{AB},
we do not spoil the BRST invariance at quantum level.

Considering now the BPHZ regularized BV master equation
(\ref{MasterB2}) and choosing conveniently for the arbitrary set of fields \break $\phi^{A_1} (x_1)  \phi^{A_2} (x_2)\,...\, \phi^{A_n} (x_n)$  just $\omega (y)$ we find:   

\begin{equation}
\alpha_1\, N_4 \,\big[ \partial_\mu (\overline\psi \gamma^\mu
\gamma_5 \psi ) \,\big]\,+\,\hbar N_4 \,\big[ {\delta^L M_1\over \delta \omega^\ast} \,\big] \,=\,0
\end{equation}

\noindent Now we can look up for a counterterm $M_1$ such that this version of the BPHZ regularized BV master equation will just  represent just the appropriate  
Axial Ward identity corresponding to eq. (\ref{BAX}), that means:

\begin{equation}
M_1 \,=\, - { e^2 \over (4\pi)^2} \,F_{\mu\nu}\,
{\tilde F}^{\mu\nu} \omega^\ast
\end{equation}

\noindent Therefore, the inclusion of this counterterm in the quantum action makes it possible to generate, by means of 
the BPHZ renormalized master equation (\ref{MasterB2}) the  Ward identity involving the divergency of the Axial current.

At this point it is important to note that if we try to incorporate the global axial anomaly in the BPHZ renormalized BV framework by simply associating a constant ghost to the global symmetry, we would not be able to reproduce the results above. 
There would be no Axial current in the action, and therefore, the Ward identities involving
the divergence of this  current would not be reproduced.
It is  the extra field $\rho$ (or, equivalently, the canonically transformed version $\omega^\ast$ ) that makes it possible to incorporate the non trivial behavior of $\partial^\mu J^5_\mu$ in the theory, through the $M_1$ term.

\section{Conclusion}
We have shown that a typical example of global anomaly: the non conservation of the axial anomaly in standard QED 
can be consistently handled within the BPHZ renormalized BV master equation.  
The whole set of Ward identities involving the divergency of the Axial current can be generated from this equation.

\vskip 1cm
\section{Acknowledgements}
We would like to thank R. Amorim , S. P. Sorella and  S. M de Souza for  important discussions. This work was partially  supported  by CNPq, FINEP, 
FUJB and CAPES
(Brazilian Research Agencies).
\vfill\eject

\end{document}